\documentstyle[preprint,aps]{revtex}

\begin{document}
\title{Electron transport properties of MgB$_{2}$ in the normal state}
\author{M.Putti, E.Galleani d'Agliano, D.Marr\`{e}, F.Napoli, M.Tassisto,}
\address{INFM/CNR, Dipartimento di Fisica, Via Dodecaneso 33, 16146 Genova, Italy}
\author{P.Manfrinetti, A.Palenzona,}
\address{INFM, Dipartimento di Chimica e Chimica Industriale, Via Dodecaneso 31,\\
16146 Genova, Italy}
\author{C.Rizzuto,}
\address{Dipartimento di Chimica e Chimica Industriale, Via Dodecaneso 31, 16146\\
Genova, Sincrotrone di Trieste, Trieste}
\author{S.Massidda}
\address{INFM Dipartimento di Fisica, Universit\`{a} di Cagliari, S.P.\\
Monserrato-Sestu Km 0.700, I-09042 MONSERRATO (CA), Italy}
\date{\today }
\maketitle
\pacs{23.23.+x, 56.65.Dy}

\begin{abstract}
We have measured the resistivity and the Seebeck coefficient of a MgB$_{2}$
sintered sample. The temperature dependence of resistivity is fitted well by
a generalized Bloch-Gr\"{u}neisen equation with a Debye temperature $\Theta
_{R}$ of 1050 K. The Seebeck coefficient is given by the sum of a diffusive
and a phonon drag term and the behavior in the temperature region $%
T_{c}<T<0.1\Theta _{R}$ follows a relationship $AT+BT^{3}$ where the two
terms are proportional to the electron and to the phonon specific heat,
respectively. The phonon drag term, here emphasized for the first time, is
rather large, indicating a strong electron-phonon interaction. The diffusive
term is positive and increases with Al doping. The comparison of the
experimental values with calculations including precise electronic structure
suggests that $\sigma $ bands give the main contribution to the Seebeck
effect. 
\end{abstract}

The discovery of \ 40 K superconductivity in MgB$_{2}$ \cite{NNMZ} has
stimulated a large discussion on the nature of the pairing and many
evidences suggest a BCS-type mechanism: the isotope effect on $T_{c}$\cite
{isotopic}, energy gap \cite{gap} and specific heat measurements \cite
{spec.heat1} and a negative pressure coefficient of $T_{c}$ \cite{Seeb1}\cite
{Seeb3}.

Electron transport properties may give insight into the normal state
conduction process and on the electronic structure, to understand whether\
MgB$_{2}$ can be considered a ''simple'' metal in which electron-phonon
interactions play the most important role, or electron correlation has to be
taken into account. From this point of view some measurements indicate that
MgB$_{2}$ behaves indeed like a metal: magnetoresistivity is large and
follows a generalized Kohler's rule \cite{kholer}, the Seebeck effect \cite
{Seeb1}\cite{Seeb3}\cite{Seeb2} is small, positive and nearly linear; on the
other hand, some Hall effect measurements \cite{Hall2} look very similar to
those of cuprates decreasing as $1/T$ with temperature. Therefore, for a
better understanding of basic properties of MgB$_{2}$, new measurements are
required possibly on high quality samples because, in sintered samples, the
presence of grain boundaries can affect the transport properties. This is
not the case of the Seebeck effect, which, as far as the scattering
processes at the grain boundaries may be considered elastic, is not affected
by granularity, as well proved in cuprates \cite{tallon}; thus, measurements
of thermoelectric power (TEP) $S$ can be an useful tool to provide
information on the electronic structure of MgB$_{2}$, as shown in ref. \cite
{Seeb3} and \cite{Seeb2}, where a dependence of $S$ on the pressure and on
Al doping was observed; these authors, however, did not consider the phonon
drag contribution, which in the case of strong electron-phonon coupling has
to be important.

In this paper we present resistivity and Seebeck effect measurements on a
sintered MgB$_{2}$ sample. The compound MgB$_{2}$ was prepared by direct
synthesis from the elements: Mg and crystalline B were well mixed together
and closed by arc welding under pure argon into outgassed Ta crucibles which
were then sealed in quartz ampoules under vacuum. The samples were slowly
heated up to 950 
${{}^\circ}$%
C and maintained at this temperature for 1 day. $X$-ray diffraction shows
values of the lattice parameters, $a=3.087(1)$, $c=3.526(1)$ \AA , and the
absence of extra reflections. The specimen for transport measurements, ($%
2\times 3\times 12$ $mm^{3}$), has been prepared by pressing the powders in
a stainless-steel die into a pellet which was then sintered by heat
treatment at 1000 
${{}^\circ}$%
C for 2 days.

The resistivity measurements were performed using a standard four probe
technique and the TEP was measured using an a.c. technique described
elsewhere \cite{putti} with sensitivity of 0.5\% and accuracy of 1.5\%. The
gradient applied to the sample was varied from $1$ to $3$ $K/cm$ and the
frequency from $\ 0.003$ to $0.008$ $Hz;$ the data were acquired with a
slowly rising temperature ($1$ $mK/sec$).

The resistivity measurements are presented in Fig. 1 up to 300 K; the
transition region is enlarged in the inset. The critical temperature defined
at half of the transition is $T_{c}=38$ $K$ with amplitude $\Delta T_{c}\sim
0.3$ $K$. The room temperature resistivity has a value of $130$ $\mu \Omega
cm$, whereas the resistivity at $T=40$ $K$ is $40$ $\mu \Omega cm$.

We fit the temperature dependence of the normal state resistivity to the
expression:

\begin{equation}
\rho (T)=\rho _{0}+\rho _{ph}(T) ,  \label{1}
\end{equation}

where $\rho _{0\text{ }}$is the temperature-independent residual resistivity
and $\rho _{ph}(T)$ the phonon-scattering contribution assumed of the
generalized Bloch-Gr\"{u}neisen form:

\begin{equation}
\rho _{ph}(T)=(m-1)\rho ^{\prime }\Theta _{R}\left( \frac{T}{\Theta _{R}}%
\right) ^{m}\int\limits_{0}^{\Theta _{R}/T}\frac{z^{m}dz}{(1-e^{-z})(e^{z}-1)%
},  \label{2}
\end{equation}

where $\Theta _{R}$ is the Debye temperature, $\rho ^{\prime }$ is the
temperature coefficient of resistivity for $T\gg \Theta _{R}$ and $m=3-5$.
Eq. (2) reduces to $\rho (T)=\rho _{0}+const\times T^{m}$ for $T\ll \Theta
_{R}$; indeed we see that from 40 to 100 K the resistivity is well fitted by
a power law $\rho (T)=\rho _{0}+const\times T^{3}$ \cite{kholer}. The best
fit to our data is obtained with $m=3$, $\rho _{0}=39.7$ $\mu \Omega cm$, $%
\Theta _{R}=1050$ $K$, $\rho ^{\prime }=0.49$ $\mu \Omega cm/K$ and is shown
in Fig.1 as a continuous line. The $\Theta _{R}$ value is in fair agreement
with $\Theta _{D}$ obtained from heat capacity measurements \cite{spec.heat2}
although lower values have also been reported \cite{spec.heat1}. $\rho
^{\prime }$ which is proportional to the dimensionless $\lambda _{tr\text{ }}
$coupling coefficient, has the same value that in A15 \cite{A15}, consistent
with a moderately strong coupling. The low temperature $T^{3}$ behavior,
common in transition metals \cite{T3} where is related to inter-band
scattering processes, suggests that also in MgB$_{2}$ the main contribution
to the electrical conductivity comes from a more mobile band, whose carriers
are scattered into empty states of another less mobile band.

The TEP measurements are shown in Fig. 2, where we see the transition around
37 K and a continuous increase from 45 K, with curvature changing from
positive to negative above 150 K.  Above the transition region the TEP
appears in excellent agreement with the literature data\cite{Seeb1}\cite
{Seeb3}\cite{Seeb2} both in value and in behavior. This contrasts with the
wide spread of resistivity data available in the literature, which indicates
a smaller influence of disorder or granularity on TEP.

The observed temperature behavior can be analyzed considering first the
diffusive

contribution to $S$ given by the Mott formula:

\begin{equation}
S_{d}=\frac{\pi ^{2}}{3}\frac{K_{B}^{2}T}{e}\frac{\sigma ^{\prime }}{\sigma }
,  \label{3}
\end{equation}

where $e$ is charge of carriers, $\sigma $ is the electrical conductivity
and $\sigma ^{\prime }=\frac{\partial }{\partial \varepsilon }\sigma
(\varepsilon )\mid _{\varepsilon _{F}}$. Note that $\varepsilon _{F}$ must
be counted upward for electrons and downward for holes. In the isotropic
case and if the relaxation time $\tau $ is independent of the energy (this
is the case for scattering with grain boundaries), $\sigma ^{\prime }/\sigma
=3/2\varepsilon _{F}$, independent of the scattering processes and equation
(3) becomes:

\begin{equation}
S_{d}=\frac{\pi ^{2}}{2e}\frac{K_{B}^{2}T}{\varepsilon _{F}},  \label{4}
\end{equation}

Second, we must consider a phonon drag term $S_{g}$, arising for
temperatures lower than the Debye temperature, when the phonon relaxation
time for interaction with other phonons and impurities is much longer than
the relaxation time for phonon-electron interactions. For pure isotropic
metals and considering only electron-phonon normal processes, an upper bound
for $S_{g}$ can be estimated to be of the order \cite{ph.drag}

\begin{equation}
S_{g}=\frac{C_{ph}}{3ne},  \label{6}
\end{equation}

where $C_{ph}$ is the phonon specific heat per unit volume. Thus, for $T\ll
\Theta _{D}$ , and assuming $S_{g}$ to really reach this upper bound, the
total TEP $S$ will take the form:

\begin{equation}
S=S_{d}+S_{g}=AT+BT^{3},  \label{7}
\end{equation}

where:

\begin{equation}
A=\frac{\pi ^{2}K_{B}^{2}}{2e}\frac{1}{\varepsilon _{F}}  \label{8}
\end{equation}

\begin{equation}
B=\frac{\beta _{3}}{3ne}=\frac{K_{B}}{e}\frac{1}{n_{a}}\frac{4\pi ^{4}}{5}%
\frac{1}{\Theta _{D}^{3}}\text{ } .  \label{9}
\end{equation}

Here $\beta _{3}=\frac{9NK_{B}}{\Theta _{D}^{3}}$ is the coefficient of the
low temperature phonon specific heat, $N$ is the number of atoms per unit
volume, and $n_{a}$ is the number of valence\ electrons which is 4/3 in MgB$%
_{2}$. Equation (6) can be compared with the experimental TEP of MgB$_{2}$
in the temperature range $T_{C}<T<0.1\Theta _{D}\sim 100K$ which overlaps
for a large amount with our measurements range.

The inset of Fig. 2 shows the ratio $S/T$ as a function of $T^{2}$ for $40$ $%
K<T<100$ $K$. The data show a linear behavior up to a $T^{2}$ value of 8000 (%
$T=90$ $K$) and then begin to bend. The best fit performed in the range $%
2000 $ $K^{2}<T^{2}<8000$ $K^{2}$ is plotted as a continuous line and the
fit parameters are $A=1.76\times 10^{-2}$ $\mu V/K^{2}$ and $B=1.26\times
10^{-6} $ $\mu V/K^{4}$.

In Fig. 2 eq. (6) with $A$ and $B$ given by the fit is plotted as a
continuous line, while the diffusive term $S_{d}=AT$ is plotted as a dashed
line. The experimental curve is well fitted by eq. (6) up to 100 K, above
which the data change curvature and tend to increase linearly, with nearly
the same slope of $S_{d}$. The experimental phonon drag term defined as $%
S-S_{d}$, plotted as a dotted line, tends to saturate above 250 K. This is
exactly what it is expected for $S_{g}$ ; in fact, increasing the
temperature, phonon-impurity and phonon-phonon processes become more
important, and the phonon drag falls, causing a peak in the TEP. In metals
as Cu, Ag, Au, Al, the phonon drag peak occurs at about $\Theta _{D}/5$ \cite
{ph.drag}, while in our case the peak is not yet reached at $\Theta _{D}/4$
giving a further evidence of the importance of the electron-phonon coupling
in MgB$_{2}$.

We therefore find that the diffusive and phonon drag terms contribute nearly
equally to the TEP. This was not recognized both in ref. \cite{Seeb3} and 
\cite{Seeb2} because only a diffusive term, even not vanishing at $T=0$, was
considered in order to fit the data.

To further verify the consistency of the model, let us try now to relate the
coefficient $A$ and $B$ with some microscopic parameters. We first start
neglecting the multi-band character of MgB$_{2}$ and we pursue in our naive
model of isotropic free electrons. Its reliability will be discussed
afterwards in the light of the band structure effects.

The coefficients $A$ and $B$ of our sample, and those extrapolated from data
in literature for pure \cite{Seeb1} and Al doped MgB$_{2}$ \cite{Seeb2}, are
summarized in table I. The Fermi energy $\varepsilon _{F}$ and the Debye
temperature $\Theta _{D}$ are obtained directly from $A$ and $B$ (eq.s (7)
and (8)): the Debye temperature of about $1400$ $K$, is only the 30\% higher
than $\Theta _{R}$ and $\varepsilon _{F}$ for pure MgB$_{2}$ is of the order
of 2 eV and becomes 1.2 eV for Al doped samples: since the Al doping raises
the Fermi level, the decreasing of $\varepsilon _{F}$ with Al doping is a
further evidence (in addition to the positive sign) that $S_{d}$ is
dominated by holes.

Let us now turn to the TEP as resulting from the band structure of MgB$_{2}$%
. Two types bands contribute to the conduction \cite{bands1} \cite{bands2}:
two $\sigma $ bands, deriving from the $p_{x,y}$ states of B and two $\pi $
bands deriving from the $p_{z}$ states. The two sets of bands have very
different dimensional character, the $\sigma $ bands being of hole-type and
nearly 2D, and the 3D $\pi $ bands mainly of electron-type. Great relevance
in the discussion of the pairing mechanism has been given to the $\sigma $
bands \cite{bands2}, and the positive sign of the Hall coefficient and of
the TEP, as well as the increasing of the latter with Al doping, confirm the
importance of these bands in the transport properties. Starting from the
precise electronic structure calculations described in Ref. \cite{bands1},
we have computed the Seebeck tensor $S_{d}$ , as a function of the chemical
potential shift, to reproduce within a rigid band scheme the Al doping.
These calculations are performed using the the scheme described, e.g., in
ref. \cite{bands1,Allen}. Because of hexagonal symmetry, the independent
components are $S_{xx}=S_{yy}$ and $S_{zz}$. Briefly, if $\Omega $ is the
unit cell volume and $v_{i}({\bf k},n)$ are the cartesian components of the
Fermi velocities for the $n-th$ band, we obtain the TEP components $S_{ii}$
as:

\begin{equation}
S_{ii}(T)= \frac{K_{B}}{e}\frac{\int d\varepsilon \frac{(\varepsilon -\mu )}{%
K_{B}T}\sigma _{ii}(\varepsilon )(-\frac{df}{d\varepsilon })}{\int
d\varepsilon \sigma _{ii}(\varepsilon )(-\frac{df}{d\varepsilon })}
\label{10}
\end{equation}

where

\begin{equation}
\sigma _{ij}(\varepsilon )=\frac{e^{2}}{\Omega }\sum_{{\bf k},n}v_{i}({\bf k}%
,n)v_{j}({\bf k},n)\tau ({\bf k},n)\delta (\varepsilon ({\bf k}%
,n)-\varepsilon )
\end{equation}

If we consider $\tau$ to be isotropic and independent of energy, it will not
affect the final result. As pointed out in Ref.\cite{Allen}, $S_{ii}$
vanishes in the approximation $-\frac{df}{d\varepsilon }=\delta (\varepsilon
-\mu )$, and it would therefore be appropriate to include the energy
dependence of $\tau $ as well. Unfortunately, it is not easy to obtain a
consistent $\tau (\varepsilon )$ approximation, and we therefore use the
most conservative, $\tau =const$, approach.

$S_{d}$ shows the expected linear behavior as a function of temperature, and
we therefore plot in Fig.3 $S_{d}/T$ , as a function of the electron doping,
in a rigid band scheme. Fig. 3(a) gives the tensor components, while to
compare with experiment we show in Fig. 3(b) the average of $S_{d}/T$ over
directions, $\overline{S}_{d}/T$ . Since both numerator and denominator in
the definition of $S_{d}/T$ contain a band summation, there is no clear cut
distinction between the $\sigma $ and $\pi $ contributions (in fact, when we
consider the $\sigma $ bands alone, the resulting $S_{d}/T$ reproduces the
free-electron results along the $x$ and $y$ directions, and is zero along $z$%
, up to the 2D to 3D crossover when $\mu $ crosses the $\sigma $ band top at 
$\Gamma $, and $S_{zz}$ grows to finally give isotropic results). With this
warning, in order to better understand our results, we decompose $S_{d}/T$
in terms of $\sigma $ and $\pi $ bands contributions (i.e. we decompose the
numerator of eq. (9)). We notice that for small values of doping the
dominant, positive contribution comes from $\sigma $ bands; when the
chemical potential goes beyond the $A$ point maximum this contribution
disappears and the resulting $S_{d}/T$ is much smaller and negative,
identical to the $\pi $ contribution. The smallness of the latter relative
to the $\sigma $ contribution can be easily understood: the numerator of eq.
(9) essentially monitors the $\sigma _{ii}(\varepsilon )$ derivative with
respect to $\varepsilon $, and this quantity is much larger for the $\sigma $
bands. As a function of the doping $x$, $S_{d}/T$ initially grows, and then
it bends down, especially when $\mu $ goes beyond the 2D to 3D crossover,
but also because of the change the $\pi $ contribution.

The calculated values are smaller, by a factor of about 1.7-2, than the
experimental values of table I. The trend as a function of doping, however,
is similar in experiment and theory; it would be interesting, in this
respect, to obtain the experimental values for larger doping. While the
agreement between theory and experiment is not quantitatively good, it may
considered to be satisfactory based on the fact that our calculations do not
include the energy dependence of $\tau $, which may result into quantitative
differences \cite{Allen}. As a speculation, we may say the following: our
understanding of MgB$_{2}$ indicates a strong electron-phonon coupling for
the $\sigma $ band top (mostly with the $E_{2g}$ phonon mode). Very likely,
such a strong coupling will depend on the energy location, relative to the $%
\sigma $ band top, resulting into an energy dependent $\tau (\varepsilon )$.
We also stress that our calculations do not contain any form of
renormalization; it has been a controversial question, in the literature\cite
{rinorm}, whether renormalization should affect the bare band-strutture
results. If this would be the case, the strong electron-phonon coupling in
MgB$_{2}$ would bring the theoretical value into agreement with experiment.
We may speculate that the presence of contributions to TEP from different
bands, having different electron-phonon couplings, should prevent a complete
cancellation of renormalization effects in the final result.

In summary, we analyzed in detail the resistivity and the thermoelectric
power in the normal state and we showed that these transport properties can
described within an independent electron framework, by taking into account
the high phonon frequencies and the strong electron-phonon coupling. In
particular the TEP is the sum of a diffusive and a phonon drag terms which
contribute nearly equally to it. The phonon drag term was not previously
recognized, in fact its peak is shifted above room temperature by the high
Debye temperature and the strong electron-phonon interaction. The diffusive
term is positive and increases with Al doping. The comparison of the
experimental values with the theoretical ones, calculated starting from the
precise electronic structure suggests that $\sigma $ bands give the main
contribution to the TEP. Further investigation will be necessary to verify
this result; in particular transport properties on samples with higher level
of Al doping will be a useful tool to better investigate the role of the $%
\sigma $ bands, whose relevance in the pairing mechanism has been strongly
advocated.

\begin{center}
{\bf Figure captions}
\end{center}

Figure 1: $\rho $ as a function of temperature; the transition region is
enlarged in the inset. The best fitting curve obtained with $\rho _{0}=39.7$ 
$\mu \Omega cm$, $\Theta _{R}$ $=1050$ $K$, $\rho ^{\prime }=4.95\times
10^{-1}$ $\mu \Omega cm/K$ is reported as a continuous line.

Figure 2: $S$ as a function of temperature; the continuous line is given by
eq. (6) with $A=1.76\times 10^{-2}$ $\mu V/K^{2}$ and $B=1.26\times 10^{-6}$ 
$\mu V/K^{4}$; the dashed line is the diffusive term $S_{d}=AT$; the dotted
line is the experimental phonon drag term defined as $S-S_{d}$. In the inset
it is shown $S/T$ as a function of $T^{2}$; the continuous line is the best
fit performed in the range $2000$ $K^{2}<T^{2}<8000$ $K^{2}$.

Figure 3: $S_{d}/T$ as a function of doping, estimated in a rigid band
scheme from the integrated DOS. (a) Tensor components and average; (b)
Average $S_{d}/T$ and its band decomposition, as explained in the text.

\bigskip

\begin{center}
{\bf Table captions}
\end{center}

Table 1: The coefficienta $A$ and $B$ obtained by fitting $S/T$ vs. $T^{2}$
(eq.(6)); $\Theta _{D}$ as obtained by eq. (8); $\varepsilon _{F}$ as
obtained by eq. (7).

\begin{center}
\bigskip {\bf Table I}

\begin{tabular}{ccccc}
& $A$ ($\mu V/K^{2})$ & $B$ ($\mu V/K^{4})$ & $\Theta _{D}$ ( $K)$ & $%
\varepsilon _{F}$ ($eV)$ \\ 
MgB$_{2}$ & 1.76$\times 10^{-2}$ & 1.26$\times $10$^{-6}$ & 1430 & 2.1 \\ 
$^{(a)}$MgB$_{2}$ & 1.70$\times 10^{-2}$ & 1.30$\times $10$^{-6}$ & 1430 & 
2.2 \\ 
$^{(b)}$MgB$_{2}$ & 2.00$\times 10^{-2}$ & 1.30$\times $10$^{-6}$ & 1430 & 
1.8 \\ 
$^{(b)}$Mg$_{0.95}$Al$_{0.05}$B$_{2}$ & 2.88$\times 10^{-2}$ & 1.41$\times $%
10$^{-6}$ & 1390 & 1.3 \\ 
$^{(b)}$Mg$_{0.9}$Al$_{0.1}$B$_{2}$ & 2.99$\times 10^{-2}$ & 1.49$\times $10$%
^{-6}$ & 1360 & 1.2
\end{tabular}
\end{center}

$\bigskip $

$^{(a)}$ref.\cite{Seeb1}; $^{(b)}$ref.\cite{Seeb2}.

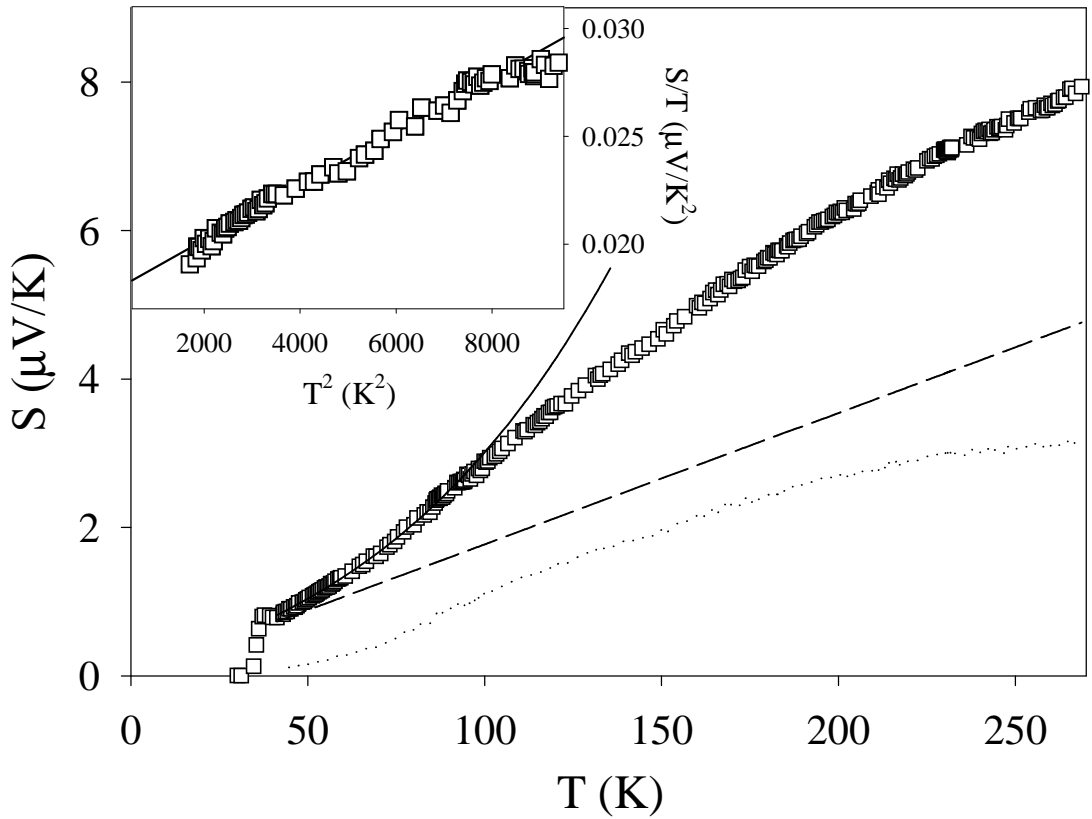
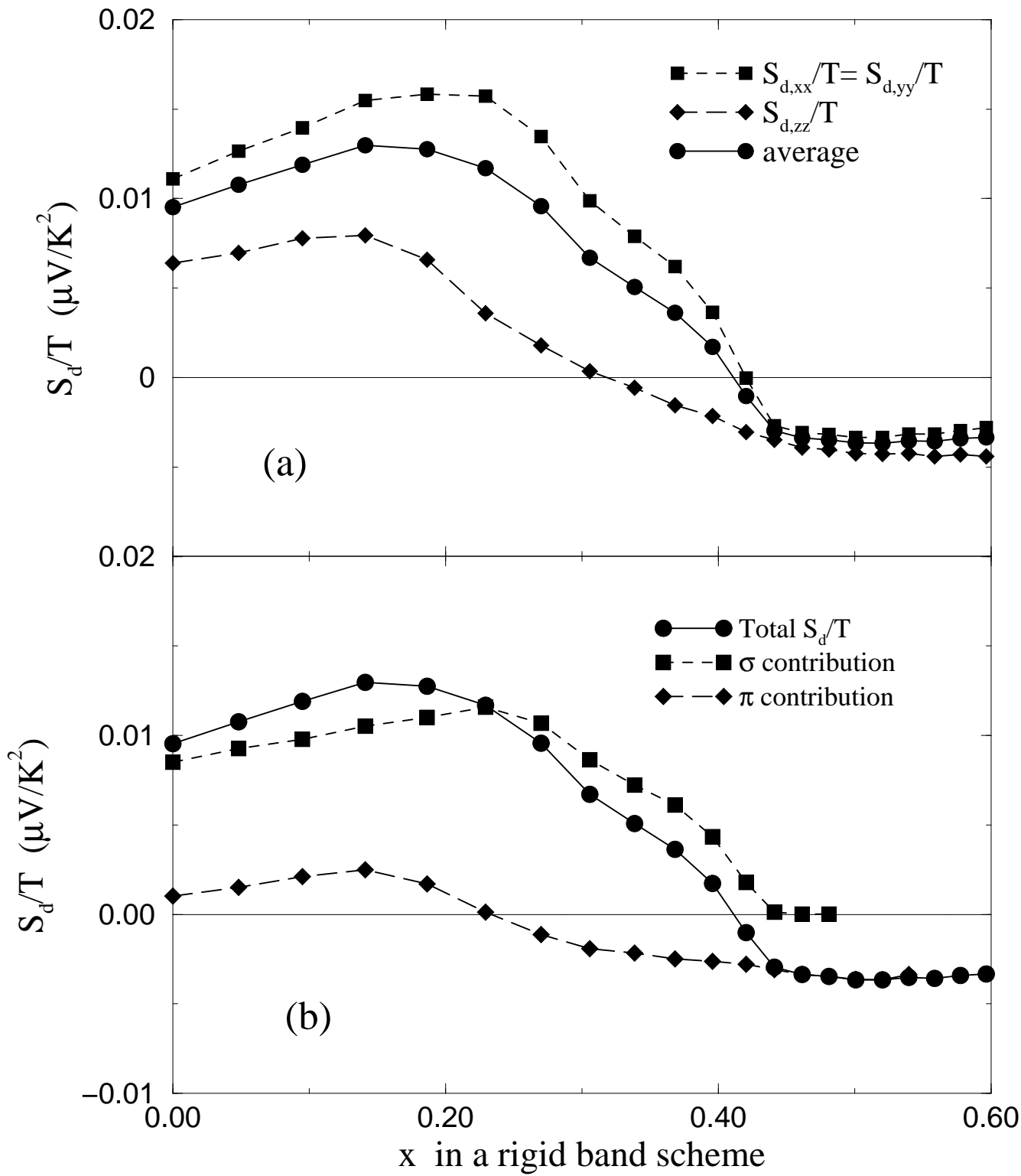
\end{document}